\documentclass[a4paper]{jpconf}
\usepackage{amsfonts,amssymb,amsmath,amsgen,amsopn,amsbsy,theorem,graphicx,epsfig}
\usepackage{graphicx}
\usepackage{color}
\usepackage{cite}
\begin{document}
\title{$\Lambda$CDM cosmology with a quiescent anisotropy in a higher dimensional steady state universe}

\author{\" Ozg\" ur Akarsu${}^{1}$, Tekin Dereli${}^{2}$ and Nihan Kat{\i}rc{\i}${}^{3}$}

\address{${}^{1}$ Department of Physics, Istanbul Technical University, Maslak 34469, Istanbul, Turkey}

\address{${}^{2}$ Department of Physics, Ko\c{c} University, Sarıyer 34450, Istanbul, Turkey}

\address{${}^{3}$ Department of Electrical and Electronics Engineering, Do\u gu\c s University, \"Umraniye 34775, Istanbul, Turkey}

\ead{akarsuo@itu.edu.tr, tdereli@ku.edu.tr, nkatirci@dogus.edu.tr}

\begin{abstract}
In this work, which follows a series of studies on the higher-dimensional steady state universe idea and prepared for Professor Tekin Dereli’s Festschrift, we show the influence of the dynamical internal (unobservable) space on the evolution of the possible anisotropy of the external (observable) space. We obtain mathematically exactly the same Friedmann equation of the standard $\Lambda$CDM model for the external space, but with some remarkable physical differences. In particular, the higher-dimensional negative cosmological constant plays the role of the four-dimensional positive cosmological constant and the expansion anisotropy, viz., the shear scalar, of the external space mimics a negative cosmological constant; it would mimic a stiff fluid when allowed on top of the standard $\Lambda$CDM model. This latter feature gives us the opportunity to manipulate the CMB quadrupole temperature fluctuation, suggesting a possible answer to the fact that its observed value is lower than that predicted by the standard $\Lambda$CDM model.
\end{abstract}

\section{Introduction}
The idea that the spacetime actually has more than four dimensions, but appears four dimensional because the extra spatial dimensions are so small that would escape local detection goes back to years Einstein's general theory of gravity was introduced in 1915. Kaluza and Klein's attempt to unify gravitation and electromagnetism in the 1920s was based on the assumption that the universe we live in is five dimensional, but since the compact fifth dimension remains small, it effectively appears four dimensional~\cite{Kaluza:1921tu,Klein:1926tv} (see Ref.~\cite{OverduinWesson97} for a review). Unifying the fundamental interactions achieved in higher dimensions in more recent years provides a strong motivation for serious consideration of this possibility; anomaly-free superstring models of all fundamental interactions require a spacetime of ten dimensions for consistency, and the M-theory in which they are embedded lives in eleven dimensional spacetime, see, e.g., Ref.~\cite{Lidsey00}. It is generally assumed that all but four of the spacetime dimensions are compactified on an unobservable internal manifold, leaving an observable $(1+3)$--dimensional spacetime. On the other hand, in the early 1980s, the dynamical (cosmological) reduction of internal dimensions to unobservable scales, that is, the idea that the external (observable) physical dimensions expand while the internal dimensions contract, began to be discussed in the cosmological context~\cite{ChodosDetweiler80,Freund82,DereliTucker83}. Since then, higher-dimensional cosmological models have been extensively studied in various approaches and contexts, and different possibilities have been explored; cosmological models where the external dimensions expand while the internal dimensions contract, remain static, expand at a much slower rate than the external dimensions, or exhibit some combinations of these possibilities, see, for instance, Refs.~\cite{Sahdev84,Tosa:1984gr,Sato84,Ishihara84,Demaret85,Okada86,Barrow86,Leon93,Yearsley96,BleyerZhuk96,RainerZhuk00,Mohammedi02,Darabi:2003st,Bringmann03,Carroll09,Reyes11,akarsuss,HoKephart10,Akarsu:2012dj,Akarsu:2012vv,Fiorini:2013hva,Akarsu:2014dxa,Akarsu:2015csa,Sloan:2016kbc,Akarsu:2017fjs,Benisty:2018gzx,Rasouli:2018owa,Kim:2018mfv,Russo:2018akp,Middleton:2019sev,Akarsu:2019oem}.

There is an interesting class in which the external and internal dimensions are dynamical but the total volume of the higher-dimensional space remains constant~\cite{Freund82,DereliTucker83,BleyerZhuk96,RainerZhuk00,HoKephart10,Akarsu:2012vv,Akarsu:2015csa}. In particular, Dereli and Tucker~\cite{DereliTucker83}, inspired by the inflationary model introduced by Guth~\cite{Guth81}, constructed a higher-dimensional general relativistic cosmological model assuming both external and internal spaces are flat, homogeneous and isotropic, but while the external space expands exponentially and the internal space contracts exponentially. It is found that the volume of the higher-dimensional space remains constant, and also that the energy density of the higher-dimensional fluid remains constant. These findings are interpreted as a leakage of mass from the contracting $n$--dimensional internal space to the expanding $3$--dimensional external space, given that the $3$--dimensional effective energy density measured in the external space also remains constant. According to this, the matter we observe is actually neither created nor exhausted, but is redistributed between the external and internal spaces. The effective $(1+3)$--dimensional universe (with an unobservable $n$--dimensional internal space) expanding exponentially adheres to the Perfect Cosmological Principle and simulates the Steady-State Theory of the Expanding Universe~\cite{BondiGold,Hoyle:1948zz} with a natural mechanism for maintaining a constant density of the matter without modifying the mathematical structure of the general theory of relativity, in contrast to the original the Steady-State Theory of the Expanding Universe, which suggests a continuous creation of matter via the introduction of the `creation field' into the Einstein field equations.

Dereli along with his collaborators has studied various extensions of the model in Ref.~\cite{DereliTucker83}, by borrowing the higher-dimensional constant volume feature as an ansatz, to explore possibility of obtaining dynamics beyond the exponential expansion, say, an effectively $(1+3)$--dimensional universe model that resembles the standard Lambda-cold dark matter ($\Lambda$CDM) model~\cite{Dodelson03}. In Ref.~\cite{akarsuss}, Akarsu and Dereli an extend it by allowing non-flat internal/external space, and explore a class of general relativistic cosmological solutions in the presence of a higher-dimensional ideal fluid that yields constant energy density along with a cosmological constant. Various dynamics for the external space were obtained; solutions with time varying deceleration parameter (for which, $q=-1$ corresponds to the exponential expansion/contraction), including oscillating and bouncing solutions when the flat/curved external and curved/flat internal spaces are considered. In Ref.~\cite{Akarsu:2012vv}, the same authors extend it by considering the dilaton gravity (inspired by the low-energy effective string theory, which already requires the presence of extra spatial dimensions~\cite{Lidsey00}) and present $(1+3+n)$--dimensional cosmological solutions with an arbitrary dilaton coupling constant $\omega$ and exponential dilaton self-interaction potentials in the string frame. In particular, to determine the cases that are consistent with the observed universe and the primordial nucleosynthesis, they focus on the class in which the external space expands with a time varying deceleration parameter and discuss the effect of the number of the internal dimensions and of the value of the dilaton coupling constant. It is found that the external space starts with a decelerated expansion rate and evolves into an accelerated expansion phase at late times tuned by the values of $\omega$ and $n$, but ends with a Big Rip~\cite{Caldwell:2003vq} in all cases. They then discuss the cosmological evolution in further detail for the case $\omega = 1$, which corresponds to the anomaly-free supertring theory for $n=6$, and to the anomaly-free bosonic string theory for $n=22$~\cite{Lidsey00}; and find results quite consistent with the observed universe we know today. In Ref.~\cite{Akarsu:2015csa}, Akarsu, Dereli, and Oflaz consider a higher-dimensional modified gravity theory with an action that includes dimensionally continued Euler-Poincar\'{e} forms up to second order (known also as Gauss-Bonnet term)~\cite{Lovelock:1971yv}, which live in the presence of extra dimensions, assuming a matter field that yields different dynamical pressures in the external and internal spaces. It is shown that the second order Euler-Poincar\'{e} term in the constructions of higher-dimensional steady state cosmologies could be crucial in finding effectively $(1+3)$--dimensional universe models resembling the standard $\Lambda$CDM model.

In the current paper, which follows the above-mentioned series of studies on the higher-dimensional steady state universe idea and prepared for Professor Tekin Dereli’s Festschrift, we briefly investigate the influence of the internal space on the evolution of the possible anisotropy of the external space. We obtain mathematically exactly the same Friedmann equation of the standard $\Lambda$CDM model for the external space, but with some interesting physical differences. In particular, the higher-dimensional negative cosmological constant plays the role of the four-dimensional positive cosmological constant and the expansion anisotropy, namely, the shear scalar, of the external space mimics a negative cosmological constant; it would mimic stiff fluid~\cite{Zeldovich:1961sbr,Barrow1978} when allowed on top of the standard $\Lambda$CDM model~\cite{GEllisBook,Akarsu:2019pwn,Akarsu:2021max}. This latter feature of the model gives us the opportunity to manipulate the cosmic microwave background (CMB) quadrupole temperature fluctuation, suggesting a possible answer to the fact that its observed value is lower than that predicted by the standard $\Lambda$CDM model~\cite{Bennett11,Schwarz:2015cma,Akrami:2019bkn}.

\section{The model}

As the theory of gravitation, we consider the extension of the conventional general theory of relativity with a cosmological constant defined in 4-dimensional spacetime to $(4+n)$--dimensional spacetime while preserving its mathematical structure;
\begin{equation}
\label{eqn:EFE}
\tilde R_{\mu\nu}-\frac{1}{2}\tilde g_{\mu\nu}\tilde R +\tilde{\Lambda}
\tilde g_{\mu\nu}=-\tilde\kappa \tilde T_{\mu\nu},
\end{equation}
where the indices $\mu$, $\nu$ run through $0,1,...,n+4$ and $0$ is reserved for the cosmic (proper) time $t$. $\tilde{R}_{\mu\nu}$, $\tilde{R}$, and $\tilde{g}_{\mu\nu}$ are, respectively, the Ricci tensor, Ricci scalar, and the metric tensor of the $(4+n)$--dimensional spacetime. Here $\tilde \kappa=8\pi \tilde G$ with $\tilde G$ being the $(4+n)$--dimensional gravitational coupling constant and $\tilde{\Lambda}$ is a $(4+n)$--dimensional cosmological constant.

 As the $(4+n)$--dimensional background geometry, we consider a spatially homogeneous but not necessarily isotropic $(1+3+n)$--dimensional spacetime metric that involves a 3--dimensional flat but not necessarily isotropic external (observable) space [described by the spatial section of the locally rotationally symmetric (LRS) Bianchi type I spacetime metric~\cite{GEllisBook}], and a flat and isotropic compact internal (unobservable) space [described by the spatial section of the spatially flat Robertson-Walker (RW) metric]:
\begin{eqnarray}
\label{eqn:metric}
dS^2=-dt^2+a(t)^2(dx^{2}+dy^{2})+ b(t)^2 dz^{2}+s(t)^2\left(d\theta_{1}^{2}+...+ d\theta_{n}^{2}\right),
\end{eqnarray}
 where $a(t)$ is the scale factor along the $x$- and $y$-axes, $b(t)$ is the scale factor along the $z$-axis of the 3--dimensional external space, and $s(t)$ is the scale factor of the $n$--dimensional internal space.
 
 As the $(4+n)$--dimensional fluid, we consider a fluid described by an anisotropic energy-momentum tensor that allows the pressures associated with the external and internal spaces to be different;
\begin{equation}
\label{eqn:EMT}
 {\tilde T}{_{\mu}}^{\nu}={\textnormal{diag}}[-\tilde\rho, \tilde p_{\rm ext},\tilde p_{\rm ext},\tilde p_{\rm ext},\tilde p_{\rm int},....],
\end{equation}
where $\tilde{\rho}=\tilde{\rho}(t)$ is the $(4+n)$--dimensional energy density, $\tilde{p}_{\rm ext}=\tilde{p}_{\rm ext}(t)$ and $\tilde{p}_{\rm int}=\tilde{p}_{\rm int}(t)$ are the pressures that are associated with the external and internal dimensions, respectively. Since we do not know the nature of the higher-dimensional physical ingredients of universe, we are allowed to consider the possibility of describing it with an energy-momentum tensor that yields distinct and dynamical pressures in the external and internal spaces.

The higher-dimensional Einstein field equations (\ref{eqn:EFE}) for the metric (\ref{eqn:metric}) in the presence of the energy-momentum tensor given in Eq.~(\ref{eqn:EMT}) lead to the following set of differential equations:
\begin{eqnarray}
\label{eqn:EFE1}
\frac {\dot{a}^2}{a^2}+{\frac{{2\dot{b}}{\dot{a}}}{ba}}+2n{\frac {{\dot{s}}\,{\dot{a}}}{sa}}+n\,{\frac {{\dot{s}}\,{\dot{b}}}{sb}}+\frac{1}{2}n(n-1){\frac{{{\dot{\it s}}}^{2}}{{s}^{2}}}+\tilde\Lambda&=&\tilde\kappa\tilde\rho,\\
\label{eqn:EFE2}
{\frac {{\ddot{\it a}}}{a}}+{\frac {{\ddot{\it b}}}{b}}+n\,{\frac {{\ddot{\it s}}}{s}
}+{\frac {{\dot{\it b}}\,{\dot{\it a}}}{ab}}+n \frac{{\dot{\it s}}}{s}\left(\frac{\dot{a}}{a}+\frac{\dot{b}}{b}\right)+\frac{1}{2} n (n-1)\,{\frac {{{\dot{\it s}}}^{2}}{{s}^{2
}}}+\tilde\Lambda&=&-\tilde\kappa \tilde p_{\rm ext},\\
\label{eqn:EFE3}
{2\frac {{\ddot{\it a}}}{a}}+n\,{\frac {{\ddot{\it s}}}{s}
}+\frac {{\dot{\it a}}^2}{a^2}+2n\,{\frac {{\dot{\it s}}\,{\dot{\it a}}}{sa}}+\frac{1}{2} n (n-1)\,{\frac {{{\dot{\it s}}}^{2}}{{s}^{2
}}}+\tilde\Lambda&=&-\tilde\kappa \tilde p_{\rm ext}, \\
\label{eqn:EFE4}
(n-1)\,{\frac {{\ddot{\it s}}}{s}}+2 (n-1)\,{\frac {{\dot{\it s}}\,{\dot{\it a}}}{sa}}+(n-1)\,{\frac {{\dot{\it s}}\,{\dot{\it b}}}{sb}}
+{\frac {{{\dot{\it s}}}^{2}}{{s}^{2}}}+2{\frac {{\ddot{\it a}}}{a}}+{\frac {{\ddot{\it b}}}{b}}+{\frac {{\dot{a}^2}}{a^2}}+{
\frac {{\dot{\it b}}\,{\dot{\it a}}}{a b}}+\tilde\Lambda&=&-\tilde\kappa\tilde p_{\rm int}.
\end{eqnarray}
The set of field equations (\ref{eqn:EFE1})--(\ref{eqn:EFE4}) should be satisfied by six  unknown functions  $a$, $b$, $s$, $\tilde\rho$, $\tilde p_{\rm ext}$, and $\tilde p_{\rm int}$, and therefore is not fully determined. We must provide two constraint equations for a full determination of the system. In this work, we are interested in the higher-dimensional cosmologies that yield constant higher-dimensional volume, and accordingly we proceed with the following ansatz, as done in Refs.~\cite{akarsuss,Akarsu:2012vv,Akarsu:2015csa};
\begin{equation}
\label{eq:volrel}
V_{3+n}=V_{\rm ext}V_{\rm int}=a^2bs^n={\rm const},
\end{equation}
where $V_{3+n}$ is the volume element of the $(3+n)$--dimensional space, while $V_{\rm ext}=a^2b$ and $V_{\rm int}=s^n$ are the volume elements of the 3--dimensional external and $n$--dimensional internal spaces, respectively.

In particular, we are interested in the evolution of the external space (the observable space, though is affected by the unobservable internal space), namely, the evolution of the average expansion rate $H_{\rm ext}$ and the shear scalar $\sigma^2_{\rm ext}$ (quantifying the expansion anisotropy) of the external space.

Therefore, we will first proceed with the definitions of these two parameters. We define the mean scale factor of the external space as $v_{\rm ext}\equiv V_{\rm ext}^{1/3}$, which in turn gives the average Hubble parameter of the external space as follows;
\begin{equation}
\label{Hxdef}
H_{\rm ext}\equiv\frac{\dot{v}_{\rm ext}}{v_{\rm ext}}=\frac{1}{3}(2H_{x}+H_{ z}),
\end{equation}
where $H_{x}=H_{y}=\frac{\dot{a}}{a}$ and $H_{z}=\frac{\dot{b}}{b}$ are the directional Hubble parameters along the $x$-- (or $y$)--axis and the $z$--axis, respectively. Similarly, the Hubble parameter of internal space reads $H_{\rm int}=\frac{{\dot{\it s}}}{s}$ from the mean scale factor of the internal space defined as $v_{\rm int}\equiv V_{\rm int}^{1/n}$. Note that the higher-dimensional constant volume condition (\ref{eq:volrel}) implies $3H_{\rm ext}+nH_{\rm int}=0$,  which in turn guarantees that an expanding external space ($H_{\rm ext}>0$) is accompanied by a contracting internal space ($H_{\rm int}<0$). This is in line with that, in contrast to the external space, the internal space must have been never expanded to observable scales or have contracted to the unobservable scales.

The measure of the anisotropic expansion in the external space section can be quantified by the shear scalar (see Ref.~\cite{GEllisBook}) defined specifically for the external space as follows;
\begin{equation}
\sigma_{\rm ext}^2\equiv\frac{1}{2}\sum_{i=1}^{3}\left(H_{i}-H_{\rm ext}\right)^{2}=  \frac{1}{3}(H_{x}-H_{z})^2.
\label{avehubble}
\end{equation}
Then, subtracting Eq.~(\ref{eqn:EFE2}) from Eq.~(\ref{eqn:EFE3}), and after some manipulations and integration, it turns out that
 \begin{eqnarray}
\label{eq:delta3hd}
\sigma_{\rm ext}^2=\frac{\Gamma}{3 \left(V_{\rm ext} V_{\rm int}\right)^2},
\end{eqnarray}
where $\Gamma\geq0$ is a non-negative definite integration constant. We notice that the shear scalar associated with the external space is controlled by the total volume element of the higher-dimensional space; i.e., its evolution is controlled by not only the external space but also the unobservable internal space. For instance, when we consider the usual 4--dimensional general relativity, in the simplest anisotropic (LRS Bianchi type I spacetime) extension of the standard $\Lambda$CDM model, the shear scalar is inversely proportional to the square of the volume element, $\sigma^2\propto V^{-2}$, see, for instance, Refs.~\cite{GEllisBook,Akarsu:2019pwn,Akarsu:2021max}. Indeed, in consistency with that, if we assume that the internal space is static, i.e., $V_{\rm int}=\rm const$, we obtain $\sigma_{\rm ext}^2\propto V_{\rm ext}^{-2}$ from Eq.~(\ref{eq:delta3hd}). However, any dynamical internal space will lead $\sigma_{\rm ext}^2$  to deviate from this behavior.

Finally, using the constant higher-dimensional volume ansatz (\ref{eq:volrel}), along with the definitions of $H_{\rm ext}$ (\ref{Hxdef}) and $\sigma^2_{\rm ext}$ (\ref{eq:delta3hd}), in the set of field equations (\ref{eqn:EFE1})--(\ref{eqn:EFE4}), we reach the following set of equations that describes our model;
\begin{align}
 3H_{\rm ext}^2 &=
 -\frac{2n}{n+3}\left(\tilde{\kappa}\tilde{\rho}+\tilde{\Lambda}+\sigma_{\rm ext}^2\right), \label{neq1}\\
 -3H_{\rm ext}^2+\frac{2n}{n+3}\dot{H}_{\rm ext}&=\frac{2n}{n+3}\left(\tilde{\kappa}\tilde{p}_{\rm ext}+\tilde{\Lambda}+\sigma_{\rm ext}^2\right),\label{neq2}\\
-\tilde\kappa\tilde p_{\rm int}&=\frac{14(n-1)}{n}H_{\rm ext}^2+\frac{3}{n}\dot{H}_{\rm ext}+\frac{1}{\sqrt{3}}H_{\rm ext}\sigma_{\rm ext}+\tilde{\Lambda}+\frac{5}{3}\sigma_{\rm ext}^2, \label{neq3}
  \end{align}
with the shear scalar of the external space is found to be constant,
  \begin{equation}
  \label{shearconst}
   \sigma_{\rm ext}^2=\rm const.    
  \end{equation}

We have now a set of field equations (\ref{neq1})-(\ref{neq3}) consisting of three equations with four unknown functions, $H_{\rm ext}$, $\tilde{\rho}$, $\tilde{p}_{\rm ext}$, and $\tilde{p}_{\rm int}$. Thus, we have freedom to introduce one more ansatz. To do so, we first recall the Friedmann equation of the standard $\Lambda$CDM model described by the usual spatially flat Friedmann-Lema\^{i}tre-Robertson-Walker (FLRW) spacetime;
\begin{equation}
\label{eq:Lcdmfried}
    3H^2=\Lambda+\kappa \rho_{\rm m0}(1+z)^3+\kappa \rho_{\rm r0}(1+z)^4,
\end{equation}
where $H$ is the Hubble parameter, $\Lambda$ is the positive cosmological constant, $\rho_{\rm m0}$ and $\rho_{\rm r0}$ are, respectively, the present-day energy densities of dust and radiation, and $z$ is the redshift; here, all defined in the usual way in 4-dimensional spacetime. Here and onward, a subscript $0$ attached to any quantity implies its value in the present-day ($z=0$) universe.

Comparing Eq.~(\ref{neq1}) and Eq.~(\ref{eq:Lcdmfried}), so as to achieve mathematically exactly the same form with the Friedmann equation of the standard $\Lambda$CDM model, assuming $\tilde\kappa>0$, we suppose the higher-dimensional energy density is negative definite ($\tilde\rho<0$) and is of the form
\begin{eqnarray}
\tilde{\rho}=-c_1(1+z)^{3}-c_2(1+z)^{4},
\end{eqnarray}
where $c_1>0$ and $c_2>0$ are constants, and $z$ is the redshift defined within the external space from $z=-1+\frac{1}{v_{\rm ext}}$ using the mean scale factor of the external space. Using this in Eq.~(\ref{neq1}), we reach the following Friedmann equation describing our model in terms of redshift;
\begin{equation}
\label{themodelfinal}
3H_{\rm ext}^2 =\frac{2n\tilde{\kappa}}{n+3}
 c_1(1+z)^{3}+\frac{2n\tilde{\kappa}}{n+3}c_2(1+z)^{4}-\frac{2n}{n+3}\tilde{\Lambda}-\frac{2n}{n+3}\sigma_{\rm ext}^2.
\end{equation}
This can be written in a more useful way, in terms of the density parameters and the Hubble constant of the external space $H_{\rm ext0}$, as follows;
\begin{equation}
\label{themodelfinal}
\frac{H_{\rm ext}^2}{H_{\rm ext0}^2} =\Omega_{\rm m0}(1+z)^{3}+\Omega_{\rm r0}(1+z)^{4}-\Omega_{\tilde\Lambda0}-\Omega_{\sigma^2_{\rm ext}0},
\end{equation}
where we defined 
\begin{equation}
\begin{aligned}
&\Omega_{\rm m0}=\frac{2nc_1}{3(n+3)H_{\rm ext0}^2} \quad,\quad \Omega_{\rm r0}=\frac{2nc_2}{3(n+3)H_{\rm ext0}^2}\\
&\Omega_{\tilde\Lambda0}=\frac{2n\tilde\Lambda}{3(n+3)H_{\rm ext0}^2}\quad \textnormal{and} \quad \Omega_{\sigma^2_{\rm ext}0}=\frac{2n\sigma_{\rm ext}^2}{3(n+3)H_{\rm ext0}^2}\geq0.
\end{aligned}
\end{equation}
First, we notice that, due to the negative definite coefficient in front of it, the shear scalar $\sigma_{\rm ext}^2$ associated with the external space contributes to the expansion rate of the external space like a negative cosmological constant. Second, the higher-dimensional cosmological constant $\tilde{\Lambda}$ should yield negative values to play the role of the four-dimensional usual positive cosmological constant in the external space and be smaller than the negative of the shear scalar associated with the external space to derive late time acceleration in the external space;
\begin{eqnarray}
\tilde{\Lambda}<-\sigma_{\rm ext}^2\leq0.
  \end{eqnarray}
 It is worth noting that, in contrast to a positive cosmological constant~\cite{Weinberg89,Sahni00}, a negative cosmological constant is not only ubiquitous in the fundamental theoretical physics without any complication, but also a theoretical sweet spot; an anti-de Sitter (AdS) background, provided by a negative cosmological constant, is welcome due to the celebrated AdS/CFT (conformal field theory) correspondence~\cite{Maldacena:1997re} and is preferred by string theory (which requires the presence of extra dimensions) and string-theory-motivated supergravities~\cite{Bousso:2000xa}. It is also interesting to observe that the shear scalar of the external space resembles a negative cosmological constant.

Note that we reproduce mathematically exactly the same form of the Friedmann equation of the standard $\Lambda$CDM model for the external space, but with an important difference that the external space is now allowed to exhibit an anisotropic expansion that yields specifically a constant shear scalar, see Eqs.~(\ref{shearconst}) and (\ref{themodelfinal}). This feature of our model, which allows late-time ellipsoidal expansion on top of the standard $\Lambda$CDM model, provides us with opportunity to manipulate the CMB quadrupole (multipole $\ell=2$ corresponding to the angular scale $\theta=\pi/2$ of the power spectrum) temperature fluctuation, $\Delta T$, with no consequences on the higher multipoles~\cite{geodif,campanelli,campa,mde,koivista}.

The evolution of the photon temperature along the $i^{\rm th}$--axis ($i=x,y,z)$ is given by $T_{i}/T_0 = e^{-\int H_i {\rm d} t} $, where $T_{0}=2.7255\pm 0.0006\,{\rm K}$~\cite{Fixsen09} is the present-day CMB monopole temperature~\cite{Barrow:1997sy}.
Accordingly, using Eq.~(\ref{avehubble}), the
difference between the photon temperatures along the $x$--(or $y$--) axis and the $z$--axis since the recombination ($z_{\rm rec}=1090$) to the present time ($z=0$) due to the anisotropic expansion, $\Delta T_{\sigma_{\rm ext}^2}\equiv T_x-T_z$, reads
\begin{equation}
\begin{aligned}
\Delta T_{\sigma_{\rm ext}^2}=T_0\int_{t_{\rm rec}}^{t_0} (H_x-H_z)\,{\rm d}t=3T_0\sqrt{\frac{n+3}{2n}\Omega_{\sigma^2_{\rm ext}0}} \int_{z=0}^{z_{\rm rec}} \frac{H_{\rm ext0}}{H_{\rm ext}}\frac{{\rm d}z}{1+z},
\label{deltaT}
\end{aligned}
\end{equation}
for small anisotropies (so $e^{-\int{H_x}dt} \simeq 1-\int H_x {\rm d}z$ etc). We use ${\rm d}t=-\frac{{\rm d}z}{H_{\rm ext}(1+z)}$ and assume that CMB was last scattered at the recombination redshift (era) $z_{\rm rec}$ ($t_{\rm rec}$). We can adopt the observational best fit values from the recent Planck release~\cite{Aghanim:2018b}; $H_{\rm ext0}=67.32\,\,{\rm km\, sec^{-1}Mpc^{-1}}$, $\Omega_{\rm m0}=0.3158$, and $\Omega_{\rm r0}\approx 10^{-4}$. The model-independent upper bounds on the present-day expansion anisotropy read $\Omega_{\sigma_{\rm ext}^20}\lesssim10^{-3}$, e.g., from type Ia Supernovae~\cite{Campanelli:2010zx,Lin:2015rza,Wang:2017ezt,Soltis:2019ryf,Zhao:2019azy,Hu:2020mzd,Kalus:2012zu}, and therefore its contribution to $H_{\rm ext}$ in Eq.~(\ref{deltaT}) can be neglected. Thus, while retaining exactly the same expansion history for the comoving volume element of the external space with that of the standard $\Lambda$CDM model, we have
\begin{equation}
\Delta T_{\sigma_{\rm ext}^2}=21\sqrt{\frac{n+3}{2n}\Omega_{\sigma^2_{\rm ext}0}}  \,\,\, K \sim 21\sqrt{\Omega_{\sigma^2_{\rm ext}0}}  \,\,\, K.
\label{deltaT2}
\end{equation}
Consequently, setting $\Omega_{\sigma_{\rm ext}^20}\simeq 5.80 \times 10^{-7}$ we obtain $\Delta T_{\sigma_{\rm ext}^2}=16\,\mu K$, which, provided that the orientation of the expansion anisotropy is set suitably, on top of the best fit standard $\Lambda$CDM model predicted value $\Delta T_{\rm std}\approx 32\, \mu K$ can be used to bring it to the observed value by the Planck satellite $\Delta T_{\rm PLK} \approx 15 \,\mu K$~\cite{Aghanim:2018b}. We would not have this opportunity in a simple LRS Bianchi type I spacetime extension of the standard $\Lambda$CDM model; in that case, $\sigma^2\propto(1+z)^6$ and the upper bound on $\Omega_{\sigma^20}$ from the big bang nuclesynthesis (BBN) light element abundances, $10^{-23}$, is one order of magnitude tighter than even the tightest upper bounds obtained from the CMB data~\cite{Barrow:1976rda,Barrow:1997sy,Saadeh:2016sak,Akarsu:2019pwn,Akarsu:2021max}.

Finally, as the $4$-dimensional effective gravitational coupling strength measured by the observers who are aware of only the external space (three spatial dimensions) is inversely proportional to the volume of the internal space, $G_4\propto\frac{1}{V_{\rm int}}$, the constant higher-dimensional volume assumption (\ref{eq:volrel}) leads to
\begin{equation}
\frac{\dot G_4 }{G_4} \propto -\frac{\dot{V}_{\rm int}}{V_{\rm int}}=\frac{\dot{V}_{\rm ext}}{V_{\rm ext}}=\,3H_{\rm ext}.
\end{equation}
Using the Planck 2018 best-fit value $H_{\rm ext0}=67.32\,\,{\rm km\, sec^{-1}Mpc^{-1}}$ we find
\begin{equation}
\frac{\dot G_4}{G_4}\bigg|_{z=0}=3H_{\rm ext0}=2.05\times 10^{-10} \,\,{\rm yr}^{-1}
\end{equation}
for the present-day value of the rate of change of $G_4$. We note that this value is consistent with the observational bounds on the possibly varying Newton's constant, e.g., $\big|\frac{\dot G}{G }\big|_{z=0} \leq 10^{-10}-10^{-12}\,\, {\rm yr}^{-1}$, see Ref.~\cite{Uzan:2010pm}. 

\section{Conclusion}

In this paper, which follows a series of studies on the higher-dimensional steady state universe idea, we have presented a brief exploration of the influence of the dynamics of the compact internal (unobservable) space on the evolution of the possible anisotropy of the external (observable) space. We have obtained mathematically exactly the same Friedmann equation of the standard $\Lambda$CDM model for the external space, but with some interesting physical differences. In particular, we have shown that the higher-dimensional negative cosmological constant plays the role of the usual four-dimensional positive cosmological constant and the shear scalar, which quantifies the expansion anisotropy, of the external space resembles a negative cosmological constant, as a consequence specific to the assumption of constant higher-dimensional volume. This latter feature of the model has given us the opportunity to manipulate the CMB quadrupole temperature fluctuation to bring its value predicted by the standard $\Lambda$CDM model to that observed by the Planck satellite. Our study also provides an example of how future precision observations that can reveal the properties of expansion anisotropy would be important in testing the higher-dimensional cosmological models with a dynamical internal space.

\section*{References}

{

}
\end{document}